\documentclass[fleqn,10pt]{wlscirep}
\usepackage[utf8]{inputenc}
\usepackage[T1]{fontenc}
\usepackage{comment}
\usepackage{amssymb}
\usepackage{xcolor}
\usepackage{graphicx}
\usepackage{booktabs, threeparttable}
\usepackage{subfig}
\usepackage{url}
\usepackage{multirow}
\usepackage{amsmath}
\usepackage{tabularx}
\usepackage{float}
\usepackage{dblfloatfix}
\usepackage{comment}
\restylefloat{table}

\title{Multi-institution encrypted medical imaging AI validation without data sharing}

\author[1,+]{Arjun Soin}
\author[3,+]{Pratik Bhatu}
\author[2]{Rohit Takhar}
\author[3]{Nishanth Chandran}
\author[3]{Divya Gupta}
\author[4]{Javier Alvarez-Valle}
\author[3,*]{Rahul Sharma}
\author[2,**]{Vidur Mahajan}
\author[1,**]{Matthew P Lungren}
\affil[1]{Stanford University, Center for Artificial Intelligence in Medicine \& Imaging, CA, United States}
\affil[2]{CARING Research, New Delhi, India}
\affil[3]{Microsoft Research, Bengaluru, India}
\affil[4]{Microsoft Research, Cambridge, United Kingdom}
\affil[*]{Corresponding author rahsha@microsoft.com}

\affil[+]{Joint first authors}
\affil[**]{Joint senior authors}

\begin{abstract}
Adoption of artificial intelligence medical imaging applications is often impeded by barriers between healthcare systems and algorithm developers given that access to both private patient data and commercial model IP is important to perform pre-deployment evaluation. This work investigates a framework for secure, privacy-preserving and AI-enabled medical imaging inference using CrypTFlow2, a state-of-the-art end-to-end compiler allowing cryptographically secure 2-party Computation (2PC) protocols between the machine learning model vendor and target patient data owner. 
A common DenseNet-121 chest x-ray diagnosis model was evaluated on multi-institutional chest radiographic imaging datasets both with and without CrypTFlow2
on two test sets spanning seven sites across the US and India, and comprising 1,149 chest x-ray images. We measure comparative AUROC performance between secure and insecure inference in multiple pathology classification tasks, and explore model output distributional shifts and resource constraints introduced by secure model inference. Secure inference with CrypTFlow2 demonstrated no significant difference in AUROC for all diagnoses, and model outputs from secure and insecure inference methods were distributionally equivalent. 
The use of CrypTFlow2 may allow off-the-shelf secure 2PC between healthcare systems and AI model vendors for medical imaging,  without changes in performance, and can facilitate scalable pre-deployment infrastructure for real-world secure model evaluation without exposure to patient data or model IP.
\end{abstract}
\begin{document}

\flushbottom
\maketitle
%
%
\thispagestyle{empty}
\section{Introduction}

While machine learning algorithms for healthcare applications are being developed at an ever increasing pace, problems associated with lack of generalisability and bias of these algorithms demand robust pre-deployment testing on real-world patient data. This dichotomy exposes a critical need for healthcare providers to have mechanisms for pre-deployment evaluation of clinical machine learning models to ensure that real-world model performance is in line with the expectations of the end-user of the model. Practical barriers exist while implementing such evaluation mechanisms particularly when access to both patient data and model intellectual property (IP) are necessary; these include legal, privacy, regulatory, data sharing, and more. The challenge arises from what is referred to as a “low trust” environment between model vendors and healthcare systems where confidentiality and compliance around exposure of patient data or model IP are governed by technical and legal mechanisms which are generally resource intensive. A similar scenario unfolds upon running Kaggle-like data challenge competitions to evaluate machine learning models by various teams because competitions provide no security to the model owner and model weights are revealed to the testing authorities.

On the technical side, healthcare providers may mandate on-premise deployment of models, which  often leads model developers to run their models as Docker containers while adopting hardening techniques such as disabling inbound protocols and ports, encrypting model weight files, disabling the root account, or compiling the model to C - to ensure that their IP is protected while deployed in the healthcare provider's premises. For healthcare systems with cloud infrastructure, providing anonymized patient data to vendors or platform providers still requires complex Business Associate Agreements (BAA) and Institutional Review Board (IRB) clearances, making the process of testing models time consuming and complex.\cite{kotsenas21} Given the large number of commercial models for varying use cases, using different data inputs and model deployment approaches, applying the technical and legal safeguards described above becomes extremely daunting particularly as a given platform service cannot offer all viable models or vendors that would be contenders for use. 

Less resource-intensive solutions to address the trust requirements include secure privacy-preserving machine learning approaches (PPML) to ensure privacy and confidentiality for both parties (data and model) while still allowing full analysis via secure inference. Recent advances in cryptography have made secure inference almost 1000x more efficient than prior state-of-the-art, thereby making it feasible for production usage.\cite{cryptflow,cryptflow2,sirnn}
With PPML, the hospital and the vendor can use 2-party secure inference (see Appendix A) to evaluate a model on unseen patient data, paving the way for the hospital to test the model, while simultaneously ensuring that the vendor learns nothing about the patient data and the hospital learns nothing about the proprietary model weights beyond the output prediction values.  

This work provides the first evaluation of 2-party secure model testing on multi-institutional medical datasets. Prior relevant work, PriMiA,\cite{primia} demonstrated the potential for PPML technology to protect the security of both sensitive patient data and proprietary machine learning models in healthcare with encrypted inference; however PPML remains largely under-explored in the field,  specifically in medical imaging applications, and work like PriMiA suffers from limitations such as weaker security guarantees, inability to use large models (as needed in medical imaging), requirement of domain expertise for implementation, poor reproducibility, and unsupported code-bases.\cite{primia} Furthermore, prior work has failed to outline absolute computation costs of encrypted inference and explore performance variations between the insecure and secure executions. 

The purpose of this work is to rigorously evaluate PPML technologies for use in medical imaging applications with real world medical imaging data and modern computer vision models. We seek to answer the following two research questions:
\begin{enumerate}
     \item  Is there a change in performance between insecure and secure model testing for clinical computer vision inference?
    \item What are the trade-offs for secure model testing with regard to computational, inference time, and communication overheads imposed by secure two-party computation (2PC)?

\end{enumerate}

\section{Results}
Secure and insecure model accuracy and classification performance comparisons for the two test sets can be found in Tables ~\ref{tab:aurocstanford} and ~\ref{tab:auroccaring}.
For all diagnoses across test sets, we observe that the accuracy (AUROC) of secure testing matches  the accuracy (AUROC) of insecure testing. Our model is also shown to be well-calibrated with significantly low and matching Brier scores across both, the secure and insecure outputs.\cite{brier} We provide these scores in Appendix E.  Model outputs from CrypTFlow2-based secure and insecure inference methods have matching distributions for all diagnoses across both test sets. 

Resources needed for insecure and secure model testing to understand the trade-off is shown in Table~\ref{tab:smto}. Each secure inference task (per image) takes 15 minutes to run and communicates 60GB data between the two machines. Since each secure inference can be done in parallel, the time taken to do secure inference on a test set can be reduced to time taken for a secure inference on a single image by using many VMs. The insecure run takes $\approx 0{\cdot}3$s per image and the size of an X-ray is $0{\cdot}3$MB.

\begin{table*}
\scalebox{0.9}{
\begin{threeparttable}
\centering
\caption{Secure model testing overheads}
\label{tab:smto}
\begin{tabular}{|l|r|r|}
\hline
\textbf{Model} & \textbf{Time (s}) & \textbf{Communication (GB)}\\
\hline
CheXpert U-Ones & 900 & 60\\
 \hline
\end{tabular}
\end{threeparttable}}
\end{table*}

\begin{table*}
\begin{threeparttable}[htp]
\centering
\caption{Insecure vs. Secure AUROC on CheXpert test set}
\label{tab:aurocstanford}
\begin{tabular}{|l|l|l|}
\hline
\multicolumn{1}{|c|}{\multirow{2}{*}{\textbf{Pathologies}}} &
  \multicolumn{2}{c|}{\textbf{AUROC scores (95\% CI$^\text{a}$)}} \\ \cline{2-3} 
\multicolumn{1}{|c|}{} & \textbf{Insecure inference}                   & \textbf{Secure inference}                   \\ \hline
No Finding             & 0.90 {[}0.87 - 0.93{]}              & 0.90 {[}0.87 - 0.93{]}              \\ \hline
Cardiomegaly           & 0.88 {[}0.84 - 0.9{]}              & 0.90 {[}0.87 - 0.93{]}              \\ \hline
Edema                  & 0.89 {[}0.85 - 0.92{]}             & 0.89 {[}0.85 - 0.92{]}             \\ \hline
Consolidation          & 0.81 {[}0.73 - 0.88{]}             & 0.82 {[}0.74 - 0.89{]}             \\ \hline
Pleural Effusion       & 0.95 {[}0.93 - 0.97{]}             & 0.95 {[}0.93 - 0.97{]}             \\ \hline
Average AUROC$^\text{b}$          & \multicolumn{1}{c|}{\textit{0.89 [0.84 - 0.92]}} & \multicolumn{1}{c|}{\textit{0.89 [0.85 - 0.93]}} \\ \hline
\end{tabular}
 \begin{tablenotes}
      \small
      \item  a. \textbf{AUROC scores with 95\% Bootstrapped Confidence Intervals} 
     \item  b. \textbf{Average AUROC reports the arithmetic mean of pathology-wise AUROC scores presented. Micro and Macro average AUROC scores are also presented in Figure 2} 
    \end{tablenotes}
\end{threeparttable}
\end{table*}

\begin{table*}
\begin{threeparttable}[H]
\centering
\caption{Insecure vs. Secure AUROC on CARING500 test set}
\label{tab:auroccaring}
\begin{tabular}{|l|l|l|}
\hline
\multicolumn{1}{|c|}{\multirow{2}{*}{\textbf{Pathologies}}} &
  \multicolumn{2}{c|}{\textbf{AUROC scores (95\% CI$^\text{a}$)}} \\ \cline{2-3} 
\multicolumn{1}{|c|}{} & \textbf{Insecure inference}                   & \textbf{Secure inference}  \\ \hline
Atelectasis            & 0.83 [0.78 - 0.87]                & 0.83 [0.78 - 0.87]                \\ \hline
Cardiomegaly           & 0.95 [0.93 - 0.97]                & 0.95 [0.93 - 0.97]                \\ \hline
Consolidation          & 0.88 [0.83 - 0.92]                & 0.88 [0.84 - 0.92]                \\ \hline
Pleural Effusion       & 0.97 [0.96 - 0.98]                & 0.97 [0.96 - 0.98]                \\ \hline
Average AUROC$^\text{b}$          & \multicolumn{1}{c|}{\textit{0.90 [0.88 - 0.94]}} & \multicolumn{1}{c|}{\textit{0.90 [0.88 - 0.94]}} \\ \hline
\end{tabular}
 \begin{tablenotes}
      \small
      \item  a. \textbf{AUROC scores with 95\% Bootstrapped Confidence Intervals} 
           \item  b. \textbf{Average AUROC reports the arithmetic mean of pathology-wise AUROC scores presented. Micro and Macro average AUROC scores are also presented in Figure 3} 
    \end{tablenotes}
\end{threeparttable}
\end{table*}


\begin{figure*}
    \subfloat[Insecure PyTorch inference.\label{fig:2acx}]{    \includegraphics[width=0.5\textwidth]{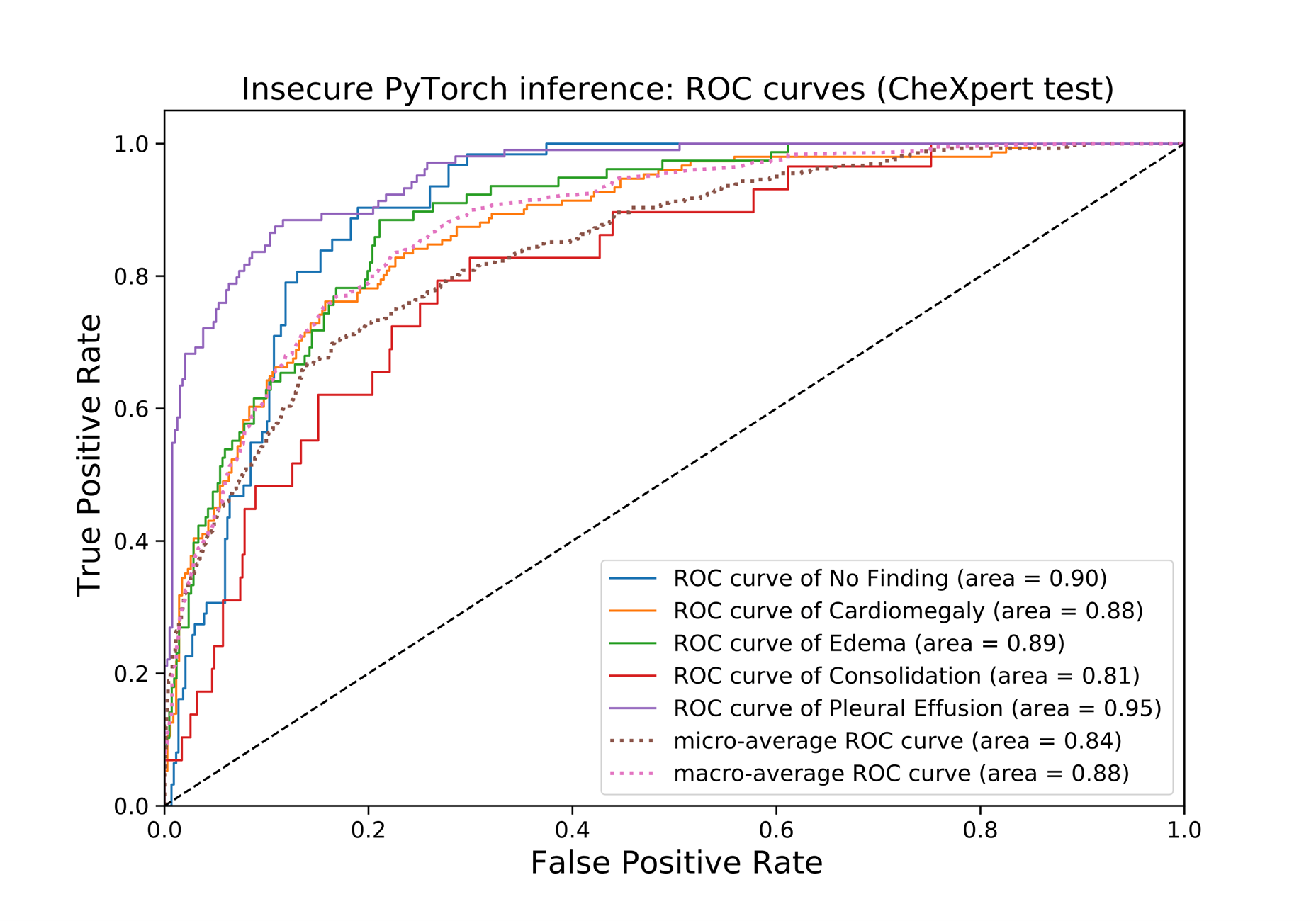}}
\hfill
    \subfloat[Secure CrypTFlow2 inference.\label{fig:2bcx}]{\includegraphics[width=0.5\textwidth]{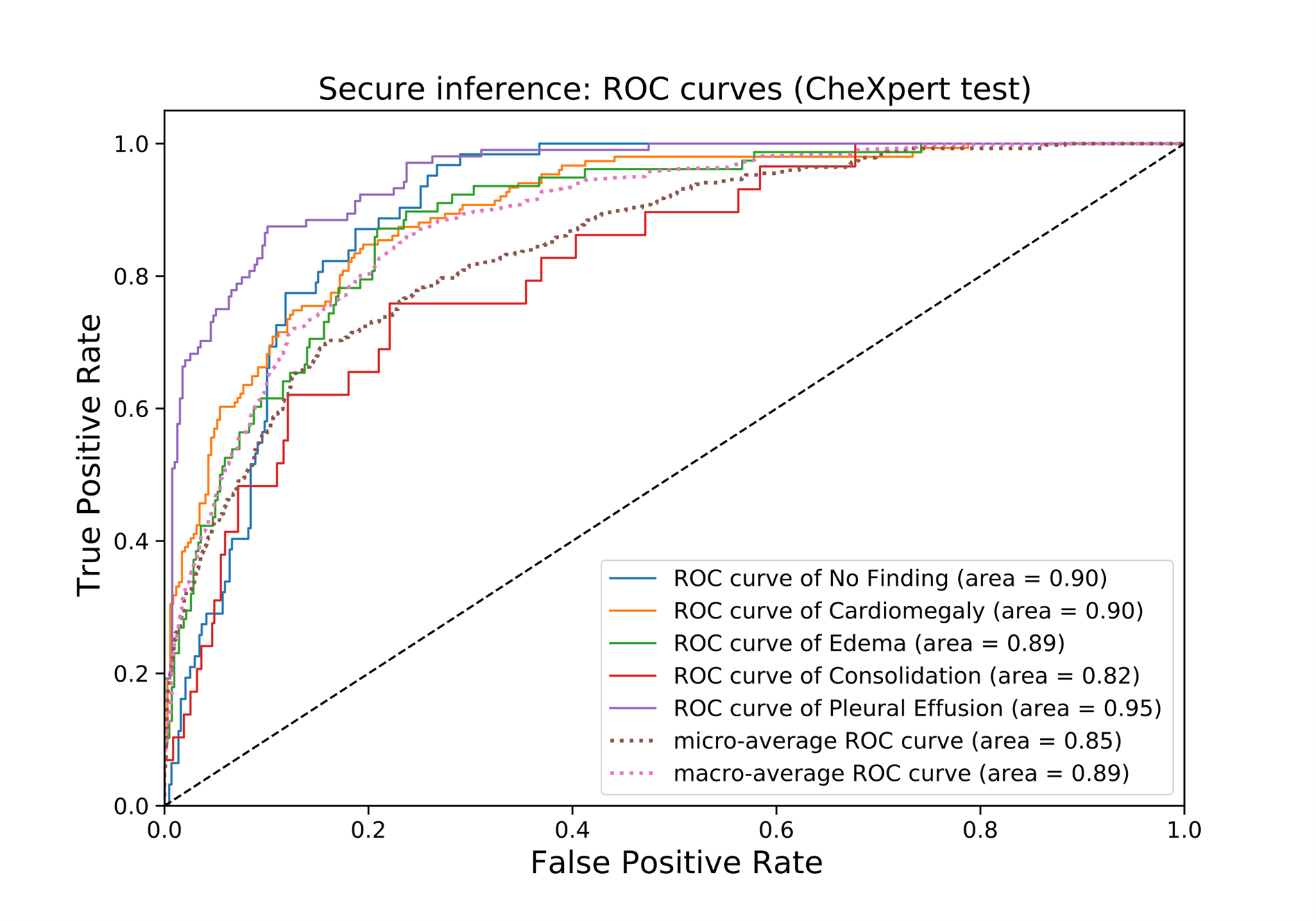}}
  \caption{Comparative model performance on CheXpert test set. Receiver operating characteristic curve (ROC) using PyTorch-based insecure inference \textbf{(a)} and CrypTFlow2-based secure inference \textbf{(b)}.
  }
  \label{fig:cx}
\end{figure*}

\begin{figure*}
    \subfloat[Insecure PyTorch inference.\label{fig:2aca}]{    \includegraphics[width=0.5\textwidth]{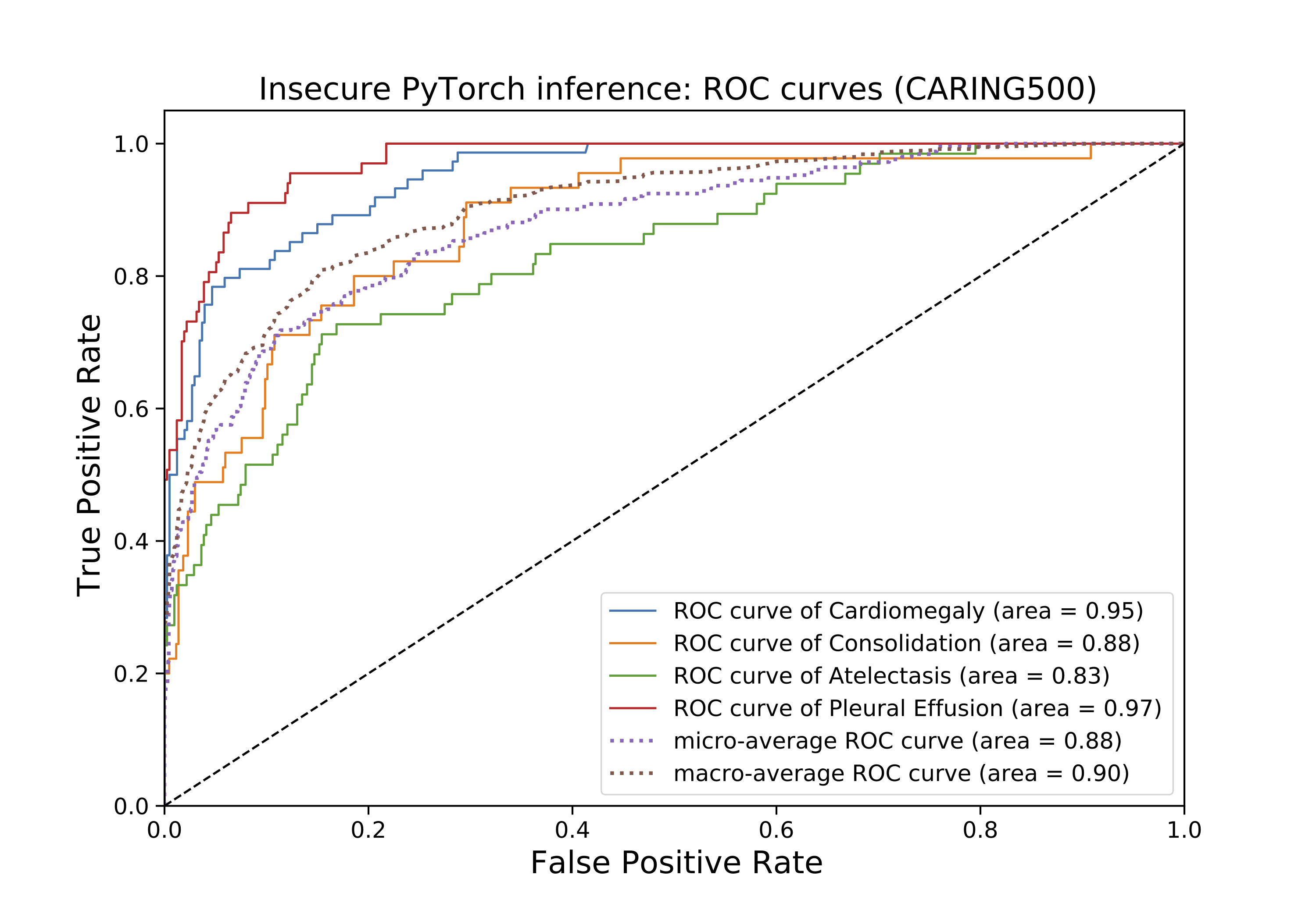}}
\hfill
    \subfloat[Secure CrypTFlow2 inference.\label{fig:2bca}]{\includegraphics[width=0.5\textwidth]{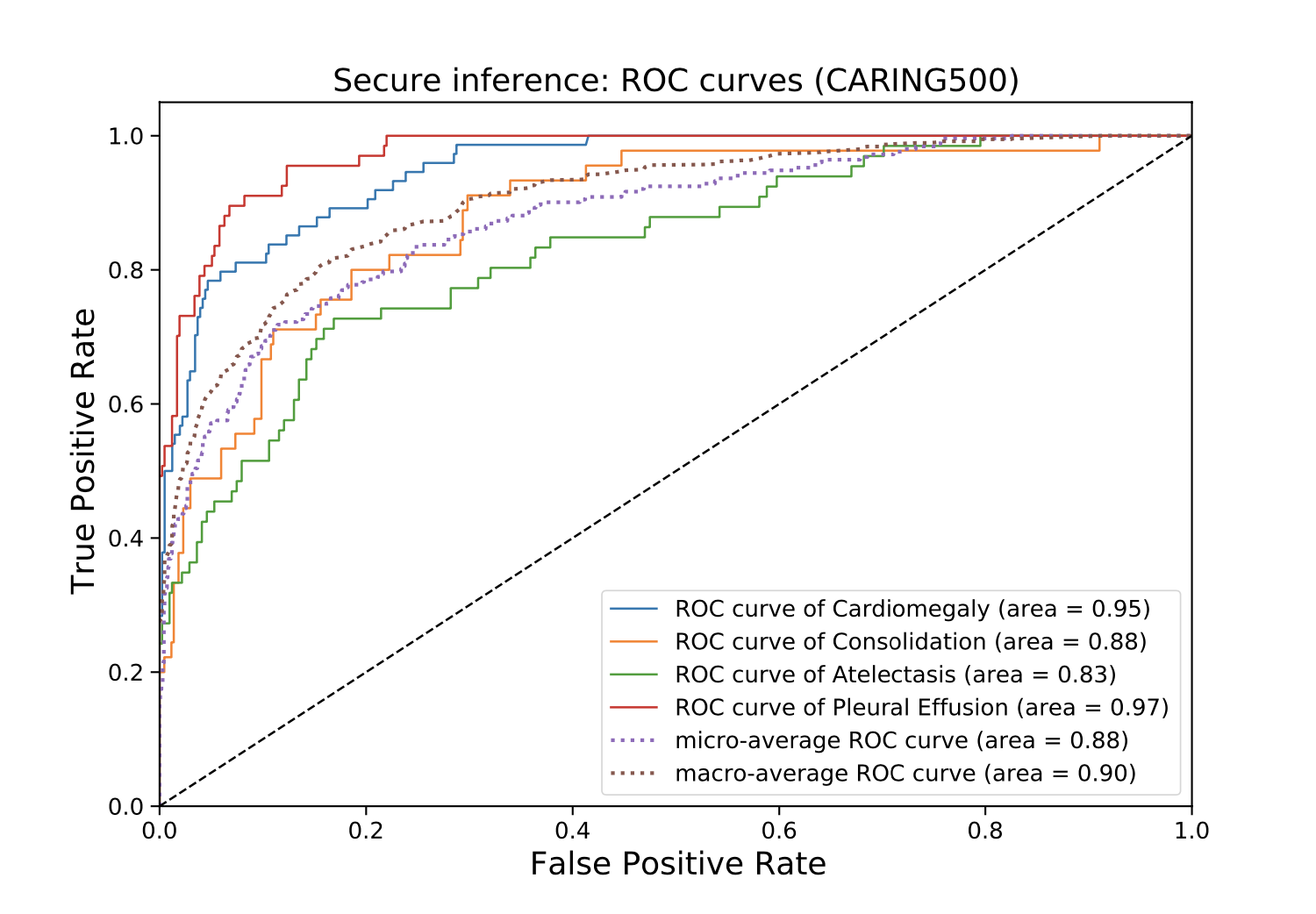}}
    \caption{Comparative model performance on CARING500 test set. Receiver operating characteristic curve (ROC) using PyTorch-based insecure inference \textbf{(a)} and CrypTFlow2-based secure inference \textbf{(b)}.
  }
  \label{fig:ca}
\end{figure*}

\begin{table*}
\centering
\begin{threeparttable}[!b]
\caption{K-S test for insecure vs. secure output distributions on CheXpert test set}
\label{tab:ksstanford}
\begin{tabular}{|l|l|l|l|}
\hline
\multicolumn{1}{|c|}{\textbf{Task}} &
  \multicolumn{1}{c|}{\textbf{Test Statistic (K-S)}} &
  \multicolumn{1}{c|}{\textbf{p-Value (K-S test)}} &
  \multicolumn{1}{c|}{\textbf{\begin{tabular}[c]{@{}c@{}}Null hypothesis \\ (secure $\equiv$ insecure)\end{tabular}}} \\ \hline
No Finding       & 0.014 & 0.99 & \textit{Accepted} \\ \hline
Cardiomegaly     & 0.076 & 0.11 & \textit{Accepted} \\ \hline
Edema            & 0.034  & 0.94 & \textit{Accepted} \\ \hline
Consolidation    & 0.058 & 0.37 & \textit{Accepted} \\ \hline
Pleural Effusion & 0.036 & 0.90 & \textit{Accepted} \\ \hline
\end{tabular}%
    \begin{tablenotes}
      \small
      \item  a. \textbf{Test performed with model output probabilities rounded off to two decimal places}
      \item  b. \textbf{Accepted hypotheses indicate matching secure and insecure model output distributions}
    \end{tablenotes}
  \end{threeparttable}
  \end{table*}

\begin{table*}
\centering
  \begin{threeparttable}[!b]
\caption{K-S test for insecure vs. secure output distributions on CARING500 test set}
\label{tab:kscaring}
\begin{tabular}{|l|l|l|l|}
\hline
\multicolumn{1}{|c|}{\textbf{Task}} & \textbf{Test Statistic (K-S)} & \textbf{p-Value (K-S test)} & 
  \multicolumn{1}{c|}{\textbf{\begin{tabular}[c]{@{}c@{}}Null hypothesis \\ (secure $\equiv$ insecure)\end{tabular}}} \\ \hline
Atelectasis      & 0.002 & 0.99    & \textit{Accepted} \\ \hline
Cardiomegaly     & 0.004   & 1.0 & \textit{Accepted} \\ \hline
Consolidation    & 0.002   & 0.99   & \textit{Accepted} \\ \hline
Pleural Effusion & 0.004  & 1.0    & \textit{Accepted} \\ \hline
\end{tabular}%
    \begin{tablenotes}
      \small
      \item  a. \textbf{Test performed with model output probabilities rounded off to two decimal places} 
      \item  b. \textbf{Accepted hypotheses indicate matching secure and insecure model output distributions}
    \end{tablenotes}
 \end{threeparttable}
\end{table*}

\section{Discussion}
The purpose of this study was to rigorously evaluate the accuracy and feasibility of PPML technology for use in medical imaging machine learning applications.  We used secure 2-party model testing for chest xray use cases in a multi-institution evaluation with readily available commodity hardware. We found no significant difference in performance compared to insecure model testing. The average performance metrics yielded were as follows: $0{\cdot}89$ AUROC (95\% CI, $0{\cdot}84$ to $0{\cdot}92$) secure, $0{\cdot}89$ AUROC (95\% CI, $0{\cdot}84$ to $0{\cdot}93$) insecure on CheXpert test set and $0{\cdot}9$ AUROC (95\% CI, $0{\cdot}88$ to $0{\cdot}94$) secure, $0{\cdot}9$ AUROC insecure (95\% CI, $0{\cdot}88$ to $0{\cdot}94$) on CARING500 test set. Moreover, we obtained distributional equivalence between secure and insecure output, with the K-S test null hypotheses being accepted for all diagnoses across both the test sets. Model calibration (as measured by the values and parity of Brier scores) was also maintained between the secure and insecure methods.



 Other systems for encrypted inference  either have not been evaluated with real-world scale models~\cite{delphi}, lack sufficient security guarantees~\cite{quantizednn}, or both.\cite{primia,securenn,ariann}
In particular, works in the latter two categories  introduce a non-colluding third party to simplify the problem of secure inference at the expense of weakening the security by introducing a dependency on a third party. Involving such a third party might prove counterproductive by increasing the legal and compliance overhead, thereby slowing down the process of testing and deployment of machine learning models in healthcare settings.
Note that 2PC is compatible with differential privacy and they can be combined for the scenarios that require masking the predicted output.\cite{abadi,pattai, ziller_medical_2021} 

CrypTFlow2 is open-source, backed by unit tests, and is actively maintained with both TensorFlow and PyTorch models, and previous preliminary work found feasibility for use with both 2D Chest X-rays and 3D CT scans.\cite{innereye,ppml} This work found that, in contrast to prior efforts,  CrypTFlow2 is an end-to-end production ready open source solution with important but manageable tradeoffs between security and inference time for real-world clinical use cases in medical imaging applications. Technologies such as CrypTFlow2 can possibly remove the requirement for de-identification and anonymisation of medical imaging data for the purposes of model inference; removing the substantial privacy-related barriers to large-scale testing of multiple models by healthcare providers without the technical hurdles of deploying models on-prem and the requirement for complicated legal safeguards will not just enable hospitals to evaluate and adopt machine learning solutions faster, but also allow for more efficient feedback loops between model users and developers. 

While insecure inference took approximately $0{\cdot}3$ seconds to process a single image, it is critical to remember that to arrive at the stage where a model deployed through insecure inference can truly be validated on real world data, countless legal agreements need to be signed, which can take several weeks, if not months, to become actionable. On the other hand, while secure inference took about 15 minutes (900 seconds) per image, the fact that each party (healthcare provider and model developer) does not have access to the other's proprietary data or IP can significantly speed up the process of executing model validation. Overall, while secure inference  takes $3000\times$ longer than insecure inference in computation time, the absolute time to validation results can be reduced dramatically. Additionally, the legal costs saved can offset the higher data transfer and compute requirements of secure inference.

There are several limitations worth highlighting. First, while this work reports results on real world clinical data with measurements of real-time inference, the validation was run retrospectively with well qualified and curated data. Real-world prospective deployment in clinical settings would give insights into the feasibility of using secure inference not just for testing but also for deployment. Second, the feasibility of running secure inference on large multi-slice and multi-parametric datasets such as CT, MRI and PET were not explored but are expected to increase time and computation. Third, a detailed costing analysis that measures the cost of computation and bandwidth required for secure inference and compares it with the legal and compliance costs of signing multiple agreements for testing AI is out of the scope of this work, but is especially important in a scenario where data sharing may become more difficult for hospitals.\cite{kotsenas21}


In conclusion, CrypTFlow2 demonstrated state-of-the-art encrypted model inference for multi-class chest radiograph classification, with performance matching native deployment.  This work, an open-source production-ready supported solution, offers a viable approach toward AI-enabled clinical imaging model inference analysis while preserving patient data and model privacy. Future work may leverage CrypTFlow2 on clinical machine learning platforms to allow for secure inference results on clinical data, affording scalable comparison across a growing variety of vendors and models, obviating the need for resource intensive legal and data sharing complexities. 

\section{Methods}

\subsection{Secure 2-party computation (2PC)} 

Secure 2-party computation (2PC)\cite{yao} considers 2 parties, usually called Alice and Bob in cryptography literature, having secret inputs $w$ and $x$ respectively. It enables Alice and Bob to securely compute any arbitrary function $F$ over $w$ and $x$. 
Furthermore, 2PC provides a strong  mathematical security guarantee that both parties learn absolutely nothing about the other party's sensitive input (beyond what can be deduced from the party's output itself). 
It is easy to see that secure model testing is a special case of 2PC for an iterative computation of $F$ on the test set, where $F$ is a machine learning model with weights $w$ and input $x$.


2PC is realized through an interactive protocol run between Alice and Bob. The parties exchange encrypted values or secret shares  with each other and further compute on them. Intuitively speaking, one can argue that each party computes on data which is cryptographically guaranteed to be indistinguishable from random  therefore revealing no information about the other party’s sensitive inputs. We provide a simplified example of a 2PC protocol in Appendix B. However, this strong security guarantee comes at a cost: computing on encrypted values or secret shares and executing the multi-round interactive protocol makes 2PC computationally more expensive to perform than insecure computations.

\begin{figure*}
    \centering
    \subfloat[Compilation of an ML model to a 2PC protocol.\label{fig:Ng1}]{
        \includegraphics[width=0.8\textwidth]{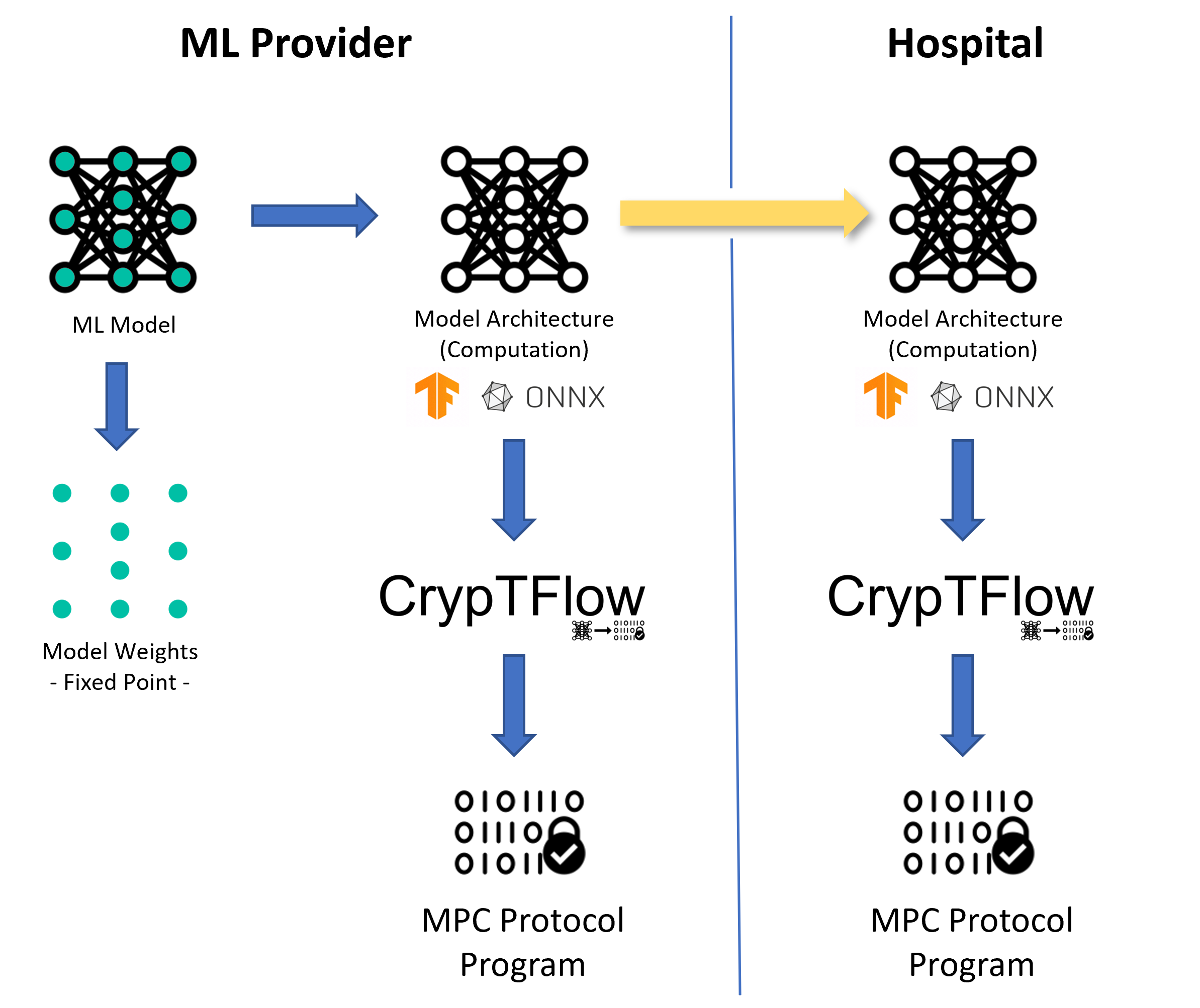}}

    \vspace{0.2in}

    \subfloat[At runtime, the parties provide secret inputs and the outputs are computed. \label{fig:Ng2}]{\includegraphics[width=\textwidth]{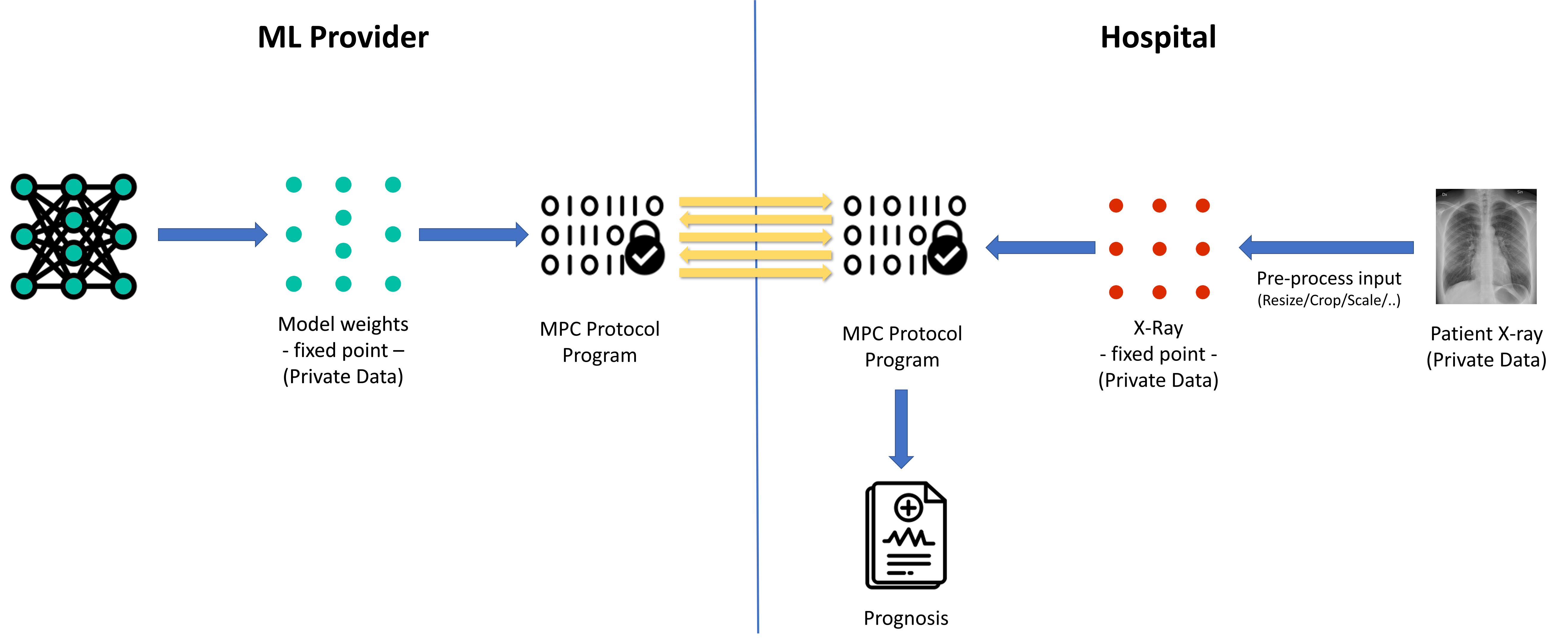}}
    \caption[Two numerical solutions]{
    
    Overview of CrypTFlow2 Compilation and Runtime Flow between ML Provider and Hospital. \textbf{a} For secure execution, we compile the ONNX model (stripped of model weights) with CrypTFlow2 to generate an executable for the ML vendor and an executable for the test set owner that are run on their respective VMs. \textbf{b} These two VMs run 2PC which ensures that sensitive data of a party is never presented in decrypted form outside its VM. The executable of the ML vendor reads model weights, the executable of the test set owners reads an image, and these two executables run a 2PC for secure inference over the network connection between them. 
    
    }
\end{figure*}

\subsection{Model use case}
Although 2PC is applicable to arbitrary ML models, we focus on the following use case: Automated chest X-ray interpretation both insecurely (using PyTorch) and securely (using CrypTFlow2). Chest X-rays are the most common imaging examination in the world, critical for the diagnosis and management of many diseases.\cite{raoof_interpretation_2012} With over 2 billion chest X-rays performed globally each year, many healthcare systems around the world lack sufficient number of trained radiologists to perform timely interpretation.\cite{mathers_projections_2006} Numerous studies have shown the competency of Convolutional Neural Networks (CNNs) in achieving performance matching radiologists  and dozens of models are available in both the research and commercial spaces.\cite{rajpurkar2017chexnet, chexpert, Wang_2017, seah21}  However, some major obstacles to clinical adoption and widespread experimentation remain data protection required to adequately protect the privacy of patients, and intellectual property considerations for AI developers to be able to preserve and protect the IP rights that they assert over their algorithms. Another noteworthy legal consideration is that of data stewardship and associated liability.\cite{gerke_ethical_2020} 


\subsubsection{Model architecture}
\label{modelarchitecture}
We use a 7-million parameter, 121-layer Dense Convolutional Network (DenseNet) trained on the CheXpert dataset.\cite{huang2018densely,chexpert} The model takes as input a single-view chest X-ray image of size 320 × 320 pixels and outputs the probability of 14 observations - \textit{No Finding, Enlarged Cardiomediastinum, Cardiomegaly, Lung Opacity, Lung Lesion, Edema, Consolidation, Pneumonia, Atelectasis, Pneumothorax, Pleural Effusion, Pleural Other, Fracture, Support Devices}. If more than one view (a frontal view and a lateral view) is available, the model outputs the maximum probability of the observations across the views. 

The model used in this work is originally implemented in PyTorch and is a submission to the CheXpert competition.\cite{chexpert} CheXpert’s competition has been running from January 2019 featuring a strong radiologist-labeled reference standard. The top 5 model checkpoints from the CheXpert competition under the \textit{U-Ones} scheme (where uncertain labels are considered positive) as of November 2020 are first selected. We then use the model checkpoint (weights) that yields the best performance scores on 5 pathologies with high prevalence in reports and clinical relevance (Cardiomegaly, Pleural Effusion, Consolidation, Edema and No Finding). Secure inference is run with this best-performing model checkpoint using CrypTFlow2 (Section~\ref{sec:eval}). 

\subsection{Test sets and metrics} To test the model on unseen patient data, we use multi-institutional chest radiographic imaging datasets on two test sets spanning seven sites across the US and India, and comprising 1,149 chest x-ray images. With strong radiologist-annotated ground truth and expert scores against which to compare algorithms, the test sets serve as strong reference standards to allow for robust evaluation of models integrated across both, our insecure and secure deep learning pipelines. Considering clinical importance and prevalence, we focus on evaluating the following pathologies for the CheXpert test set: \textit{Cardiomegaly, Pleural Effusion, Consolidation, Edema and No Finding}. Based on corresponding clinical relevance, the CARING500 test set is evaluated for \textit{Atelectasis, Cardiomegaly, Consolidation and Pleural Effusion}. The accuracy on both test sets is measured as the standard AUROC score.

\subsubsection{CheXpert test set} The CheXpert test set comprises 500 studies with 668 images from 500 patients. Eight board-certified radiologists individually annotated each study in the test set, classifying each of the 14 observations mentioned in section~\ref{modelarchitecture} into either present, uncertain likely, uncertain unlikely, or absent. The radiologist annotations were binarized such that every present and uncertain likely case is treated as positive and every absent and uncertain unlikely case is treated as negative. The majority vote of 5 radiologist annotations serves as a strong ground truth. 

\subsubsection{CARING500  test set}  This test set comprises of 481 chest X-rays obtained from six sites – three hospitals and three out-patient imaging centres – annotated by 3 radiologists at image level for the presence of ten findings -  \textit{Cardiomegaly, Fibrosis, Mediastinal Widening, Consolidation, Nodule, Pleural Effusion, Pneumothorax, Calcification, Atelectasis, Pneumoperitoneum and Overall Abnormal}. Consensus of the 3 readers was taken as ground truth for testing. The data is publicly available for download (without annotations) from \url{platform.carpl.ai}.  

\subsection{Evaluation setup} 
Both the model owner (ML vendor) and the test set owner (hospital) are virtual machines (VMs) with commodity class hardware: \textit{$3{\cdot}7$ GHz Intel Xeon processor} with 4 cores and 16 GBs of RAM. The ping latency is $0{\cdot}3$ms and the network bandwidth between the two machines is 3Gbps. Such VMs can readily be obtained from  cloud service providers, e.g. Microsoft Azure.  

\subsubsection{Evaluation process} 
\label{sec:eval}
For insecure execution, we run the PyTorch CheXpert model directly on a given test set. For secure execution, we export the PyTorch model to an ONNX model, compile the ONNX model (stripped of model weights) with CrypTFlow2 to generate an executable for the ML vendor and an executable for the test set owner that are run on their respective VMs. (See Appendix C for details about PyTorch and ONNX.) These two VMs run 2PC which ensures that sensitive data of a party is never presented in decrypted form outside its VM. In particular, the executable of the ML vendor reads model weights, the executable of the test set owners reads an image, and these two executables run a 2PC for secure inference over the network connection between them. We run intermediary tests to confirm that the compiled ONNX model uses an executable by the test set owner that is devoid of any proprietary weights from the ML vendor. The ONNX computation graph initializer, that is otherwise used to recover information about model weights, yields an empty value for each experiment we run, signifying a complete absence of proprietary weights in the model received by the test set owner. A detailed flow of compilation and runtime for our experiments is outlined in Figure 3.

By using ONNX (Open Neural Network Exchange) as a frontend,\cite{onnx} CrypTFlow2 is applicable to models expressed in various ML platforms like Keras,\cite{chollet2015keras} TensorFlow,\cite{tensorflow} PyTorch,\cite{pytorch} etc.  It first converts the models to ONNX using existing converters and then compiles the ONNX model to 2PC protocols. Our model is converted using the in-built ONNX converter from the Torch library (see Appendix D). 

\subsubsection{Evaluation metrics} 
We evaluate the average time and data communication needed for secure inference on each image and compare it to the time and communication needed for the insecure inference.

The performance or accuracy on these tests is measured as the standard AUROC score. 
It is a discrimination measure which demonstrates how well the model can classify patients in two groups: those with and those without a given pathology of interest. 
The AUROC is calculated as the area underneath a curve that measures the tradeoff between true positive rate (TPR) and false positive rate (FPR) at different decision thresholds. 
A test with no better accuracy than chance has an AUROC of 0.5, and a test with perfect accuracy has an AUROC of 1.0. Accordingly, an AUROC = 0.90 indicates that 90\% of the time we draw a test X-ray from the disease group and non-disease group, the output value from the disease group will be greater.\cite{auroc} We use nonparametric bootstrap to estimate 95\% confidence intervals for the AUROC scores.
\par
To confirm that model output results from secure and insecure inference are distributionally equivalent, a two-sample Kolmogorov-Smirnov (K-S) test is calculated (Tables ~\ref{tab:ksstanford} and ~\ref{tab:kscaring} ). The K-S test is a nonparametric test of the equality of continuous, one-dimensional probability distributions that can be used to compare a sample with a reference probability distribution.\cite{kolmogorov_1951} We use inference results from insecure PyTorch inference as reference data and results from CrypTFlow2-based secure inference as the sample distribution to compare with. Under the null hypothesis, there is no difference between the two (secure and insecure) distributions. If the K-S statistic is small or the p-value is high (> 5\%), we do not reject the hypothesis that the distributions of secure and insecure model outputs are sampled from identical populations. Conversely, we reject the null hypothesis if the p-value is low (< 5\%) and conclude that CrypTFlow2 introduces distributional shift in model outputs.

\subsubsection{Data availability}
CrypTFlow2 software is publicly available under MIT license at \url{https://github.com/mpc-msri/EzPC}. To run the CheXpert model on an X-ray image securely, please follow the step-by-step instructions at
\url{https://github.com/mpc-msri/EzPC/tree/onnx-demo/Athos/onnx-demo}. CheXpert test set is not available for public use since it is reserved as a private test set. The CARING500 dataset can be accessed at \url{platform.carpl.ai} (ground truth labels are available upon request).

\appendix

\section{Secure 2-party inference}
\label{sec:si}
In the problem of secure inference there are two parties, say a patient Bob and a Vendor Alice. Alice has a pre-trained machine learning model $F$ whose weights $w$ are sensitive and can’t be revealed to anyone else. Examples of $F$ include ResNet50 on $224\times224$ images, DenseNet121 on $320\times320$ images, etc. Bob has a sensitive input $x$, e.g., a chest X-ray image, that can’t be revealed to Alice. In secure inference, the goal is for Bob to learn the output prediction of the model on his input $x$, i.e., $y=F(w,x)$. Additionally, Alice should learn nothing about Bob’s image $x$, and Bob should learn nothing about Alice’s weights $w$ beyond what can be deduced from $y$. Note that here Alice and Bob trust no third party and are mutually distrustful of each other. The problem defined here is that of 2-party secure inference between Alice and Bob. 
Furthermore, $F$ need not be restricted to be an inference algorithm and could even be a segmentation algorithm.\cite{innereye,ppml}

\section{Secure 2-party computation}
\label{app:he2pc}
A very simple example of a secure 2PC protocol now follows. Say, Alice has two values $w=(w_1,w_2)$ and Bob has $x=(x_1,x_2)$ and $F(w,x)=w_1 x_1+w_2 x_2$. Then, 2PC can be implemented using Homomorphic encryption which allows one to compute functions over encrypted values to obtain encrypted outputs. Encryption schemes supporting certain specific functions (e.g. additions and multiplications) are practical today and can be securely combined with other cryptographic techniques in order to obtain secure 2PC protocols for arbitrary functions.~\cite{cryptflow2} 

First, Bob encrypts $(x_1,x_2)$ into $(E(x_1),E(x_2))$ and sends the encryptions to Alice. Then, Alice computes $e=w_1*_{E} E(x_1)+_{E} w_2 *_{E} E(x_2)$ and returns $e$ to Bob. Bob then decrypts $e$ to get $w_1 x_1+w_2 x_2$. It is easy to see that since Alice only obtained encryptions of $x$, she doesn’t learn anything about $x$, i.e., given $(E(x_1),E(x_2))$, she cannot guess any bit of $x$ (other than guessing it at random). Note that the encrypted versions of $x_1 $ and $x_2$, that is, $(E(x_1),E(x_2))$, is of a larger size than $(x_1,x_2)$ which imposes a communication overhead. Furthermore, the homomorphic arithmetic operators $(*_E,+_E)$ are computationally more expensive than addition and multiplication of unencrypted numbers and impose a computational overhead. Hence, encrypted inference has a higher cost than insecure inference.

\section{ML frameworks background}
\label{sec:frameworksappendix}
PyTorch is an open-source machine learning (ML) framework based on the Python programming language and the Torch library.\cite{pytorch} PyTorch computes using tensors that are accelerated by graphics processing units (GPU). Tensors are arrays, multidimensional data structures that can be operated on and manipulated with APIs.

ONNX (Open Neural Network Exchange) is a format that helps with interoperability in deep learning.\cite{onnx} There are production quality converters that take ML models written in different formats and convert them to ONNX. 

\section{Intermediary model conversions}
\label{sec:modelconversionsappendix}
We convert the DenseNet-121 used for our experiments using the in-built ONNX converter from the Torch library. We verify the correctness of computation graph conversion by also converting using a PyTorch to Keras to ONNX methodology. Here, the model is first converted from PyTorch to Keras using \textit{pytorch2keras}. We then use this intermediary Keras model to get a Tensorflow Graph which is in turn converted to ONNX using \textit{tensorflow-onnx}. 

\section{Model calibration and Brier scores}
\label{sec:brierappendix}
Brier scores are used to calculate a model's accuracy of predictions, and give them a score between 0 and 1. A score of 0 indicates that a model predicted 100\% certainty for every disease class and got them all correct; a score of 1 reflects a model that predicts 100\% certainty for every disease class but is wrong in all the predictions.\cite{brier} Accordingly, we compute brier scores for both, CrypTFlow2-secure and insecure inference on both our datasets. Tables 6 and 7 present the brier score metric. We find that for all diagnoses across our multi-institutional test sets, the insecure and secure models are not only both well-calibrated (the highest recorded brier score for any disease class is less than 0.17) but also demonstrate matching calibration score values. This provides yet another additional performance-based confirmation of CrypTFlow2's validity to AI-enabled medical imaging inference. 

\begin{center}
\small
\begin{threeparttable}[H]
\centering
\caption{Brier score insecure vs. secure on CheXpert test set}
\label{tab:brierstanford}
\begin{tabular}{|l|l|l|}
\hline
\multicolumn{1}{|c|}{\multirow{2}{*}{\textbf{Pathologies}}} &
  \multicolumn{2}{c|}{\textbf{Brier scores}} \\ \cline{2-3} 
\multicolumn{1}{|c|}{} & \textbf{Insecure inference*}                   & \textbf{Secure inference*}                  \\ \hline
No Finding*             &      0.081        &     0.081         \\ \hline
Cardiomegaly*           & 0.168              & 0.168             \\ \hline
Edema*                  & 0.103             & 0.103             \\ \hline
Consolidation*          & 0.061            & 0.061             \\ \hline
Pleural Effusion*       & 0.01                & 0.01             \\ \hline
\end{tabular}
    \begin{tablenotes}
      \small
      \item  *\textbf{Scores rounded off to three decimal places}
    \end{tablenotes}
\end{threeparttable}
\end{center}

  \begin{center}
  \small
\begin{threeparttable}[H]
\centering
\caption{Brier score insecure vs. secure on CARING500 test set}
\label{tab:brierccaring}
\begin{tabular}{|l|l|l|}
\hline
\multicolumn{1}{|c|}{\multirow{2}{*}{\textbf{Pathologies}}} &
  \multicolumn{2}{c|}{\textbf{Brier scores}} \\ \cline{2-3} 
\multicolumn{1}{|c|}{} & \textbf{Insecure inference*}                   & \textbf{Secure inference*}  \\ \hline
Atelectasis*            & 0.088             & 0.088            \\ \hline
Cardiomegaly*           & 0.086               & 0.086               \\ \hline
Consolidation*          & 0.065                & 0.065              \\ \hline
Pleural Effusion*       & 0.053                & 0.053               \\ 
\hline
\end{tabular}
    \begin{tablenotes}
      \small
      \item  *\textbf{Scores rounded off to three decimal places}
    \end{tablenotes}
\end{threeparttable}
\end{center}


\end{document}